\documentclass[twocolumn,prb,showpacs,superscriptaddress,amsmath,amssymb]{revtex4}

\usepackage{graphicx}
\usepackage[english,german]{babel}
\usepackage{amsmath}
\usepackage{amssymb}
\usepackage[latin1]{inputenc}

\begin{document}

\title{Lithium Ion Transport Mechanism in Ternary Polymer Electrolyte-Ionic Liquid Mixtures -- A Molecular Dynamics Simulation Study}

\author{Diddo Diddens}
\email{d.diddens@uni-muenster.de}
\affiliation{Institut f\"ur physikalische Chemie, Westf\"alische Wilhelms-Universit\"at M\"unster, Corrensstrasse 28/30, 48149 M\"unster, Germany}
\affiliation{NRW Graduate School of Chemistry, Corrensstrasse 36, 48149 M\"unster, Germany}

\author{Andreas Heuer}
\affiliation{Institut f\"ur physikalische Chemie, Westf\"alische Wilhelms-Universit\"at M\"unster, Corrensstrasse 28/30, 48149 M\"unster, Germany}
\affiliation{NRW Graduate School of Chemistry, Corrensstrasse 36, 48149 M\"unster, Germany}

\selectlanguage{english}

\date{\today}

\newcommand{\eg}{\mbox{e.\,g.\,}\ }
\newcommand{\ie}{\mbox{i.\,e.\,}\ }
\newcommand{\etal}{{\it et al.\,}}

\newcommand{\Li}{{$\text{Li}^\text{+}$}}

\newcommand{\PnoIL}{{$\text{PEO}_{20}\text{LiTFSI}$}}
\newcommand{\PsomeIL}{{$\text{PEO}_{20}\text{LiTFSI}\cdot0.66\text{~PYR$_\mathrm{13}$TFSI}$}}
\newcommand{\PmoreIL}{{$\text{PEO}_{20}\text{LiTFSI}\cdot3.24\text{~PYR$_\mathrm{13}$TFSI}$}}
\newcommand{\PxIL}{{$\text{PEO}_{20}\text{LiTFSI}\cdot\,x\text{~PYR$_\mathrm{13}$TFSI}$}}

\newcommand{\Ptwenty}{{$\text{PEO}_{20}\text{LiTFSI}$}}
\newcommand{\Psixteen}{{$\text{PEO}_{16}\text{LiTFSI}\cdot0.556\text{~PYR$_\mathrm{13}$TFSI}$}}
\newcommand{\Ptwelve}{{$\text{PEO}_{12}\text{LiTFSI}\cdot1.111\text{~PYR$_\mathrm{13}$TFSI}$}}
\newcommand{\Peight}{{$\text{PEO}_{8}\text{LiTFSI}\cdot1.667\text{~PYR$_\mathrm{13}$TFSI}$}}
\newcommand{\Pzero}{{$\text{LiTFSI}\cdot2.815\text{~PYR$_\mathrm{13}$TFSI}$}}
\newcommand{\PminusxIL}{{$\text{PEO}_{20-\alpha x}\text{LiTFSI}\cdot\,x\text{~PYR$_\mathrm{13}$TFSI}$}}

\newcommand{\Pno}{{$\text{P}_{20}\text{S}$}}
\newcommand{\Psome}{{$\text{P}_{20}\text{S}\cdot0.66\text{~IL}$}}
\newcommand{\Pmore}{{$\text{P}_{20}\text{S}\cdot3.24\text{~IL}$}}
\newcommand{\Px}{{$\text{P}_{20}\text{S}\cdot\,x\text{~IL}$}}

\newcommand{\Ptwen}{{$\text{P}_{20}\text{S}$}}
\newcommand{\Psixt}{{$\text{P}_{16}\text{S}\cdot0.556\text{~IL}$}}
\newcommand{\Ptwel}{{$\text{P}_{12}\text{S}\cdot1.111\text{~IL}$}}
\newcommand{\Pei}{{$\text{P}_{8}\text{S}\cdot1.667\text{~IL}$}}
\newcommand{\Pz}{{$\text{P}_{0}\text{S}\cdot2.815\text{~IL}$}}
\newcommand{\Pminusx}{{$\text{P}_{20-\alpha x}\text{S}\cdot\,x\text{~IL}$}}

\newcommand{\Ptwentyonezero}{{$\text{PEO}_{20}\text{LiTFSI}\cdot0.0\text{~PYR$_\mathrm{14}$TFSI}$}}
\newcommand{\Ptwentyoneone}{{$\text{PEO}_{20}\text{LiTFSI}\cdot1.0\text{~PYR$_\mathrm{14}$TFSI}$}}
\newcommand{\Ptwentyonetwo}{{$\text{PEO}_{20}\text{LiTFSI}\cdot2.0\text{~PYR$_\mathrm{14}$TFSI}$}}
\newcommand{\Ptwentyonefour}{{$\text{PEO}_{20}\text{LiTFSI}\cdot4.0\text{~PYR$_\mathrm{14}$TFSI}$}}
\newcommand{\Ptenonezero}{{$\text{PEO}_{10}\text{LiTFSI}\cdot0.0\text{~PYR$_\mathrm{14}$TFSI}$}}
\newcommand{\Ptenoneone}{{$\text{PEO}_{10}\text{LiTFSI}\cdot1.0\text{~PYR$_\mathrm{14}$TFSI}$}}
\newcommand{\Ptenonetwo}{{$\text{PEO}_{10}\text{LiTFSI}\cdot2.0\text{~PYR$_\mathrm{14}$TFSI}$}}
\newcommand{\Pfiveoneone}{{$\text{PEO}_{5}\text{LiTFSI}\cdot1.0\text{~PYR$_\mathrm{14}$TFSI}$}}

\newcommand{\Ptwonezero}{{$\text{P}_{20}\text{S}$}}
\newcommand{\Ptwoneone}{{$\text{P}_{20}\text{S}\cdot1.0\text{~IL}$}}
\newcommand{\Ptwonetwo}{{$\text{P}_{20}\text{S}\cdot2.0\text{~IL}$}}
\newcommand{\Ptwonefour}{{$\text{P}_{20}\text{S}\cdot4.0\text{~IL}$}}
\newcommand{\Ptonezero}{{$\text{P}_{10}\text{S}\cdot0.0\text{~IL}$}}
\newcommand{\Ptoneone}{{$\text{P}_{10}\text{S}\cdot1.0\text{~IL}$}}
\newcommand{\Ptonetwo}{{$\text{P}_{10}\text{S}\cdot2.0\text{~IL}$}}
\newcommand{\Pfoneone}{{$\text{P}_{5}\text{S}\cdot1.0\text{~IL}$}}

\hyphenation{mppy}
\hyphenation{mppyTFSI}
\hyphenation{LiTFSI}
\hyphenation{propyl-pyrrol-idinium}
\hyphenation{AMBER}
\hyphenation{sander}
\hyphenation{SHAKE}

\begin{abstract}
The lithium transport mechanism in ternary polymer electrolytes, consisting 
of \Ptwenty\ and various fractions of the ionic liquid PYR$_\mathrm{13}$TFSI, is 
investigated by means of MD simulations. 
This is motivated by recent experimental findings~\cite{PasseriniJoost}, which 
demonstrated that these materials display an enhanced lithium mobility relative to 
their binary counterpart \Ptwenty. 
In order to grasp the underlying microscopic scenario giving rise to these observations, 
we employ an analytical, Rouse-based cation transport model~\cite{MaitraHeuerPRL2007}, 
which has originally been devised for conventional polymer electrolytes. 
This model describes the cation transport via three different mechanisms, 
each characterized by an individual time scale. 
It turns out that also in the ternary electrolytes essentially all lithium ions are 
coordinated by PEO chains, 
thus ruling out a transport mechanism enhanced by the presence of ionic-liquid molecules. 
Rather, the plasticizing effect of the ionic liquid contributes to the increased lithium 
mobility by enhancing the dynamics of the PEO chains and consequently also the motion 
of the attached ions. 
Additional focus is laid on the prediction of lithium diffusion coefficients from the 
simulation data for various chain lengths and the comparison with experimental data, 
thus demonstrating the broad applicability of our approach. 
\end{abstract}



\maketitle

\section{Motivation}

Solid polymer electrolytes (SPEs) are promising candidates for lithium ion batteries, 
as they are ideal to create small and light-weighted but powerful energy 
storages~\cite{Gray,BruceJCSFaraday1993}. 
The classical SPEs consist of an amorphous polymer matrix, 
\eg poly(ethylene oxide) (PEO), and a lithium salt dissolved in it~\cite{Wright,Armand}. 
By using lithium salts with large anions such as lithium-bis\-(tri\-fluoro\-meth\-ane)sulfon\-imide 
(LiTFSI), the crystallization can be suppressed as the negative charge is delocalized over the 
whole anion. 
However, at ambient temperatures, the conductivity of most SPEs is still too low for an 
efficient technological use. 
Among several other remedies~\cite{ScrosatiSSI1992,BorghiniElectrochimActa1996,BandaraElectrochimActa1998,KimSSI2002}, 
the incorporation of a room temperature ionic liquid (IL) seems to be a very promising approach~\cite{PasseriniElectrochemCommun2003,PasseriniJoost}, 
as the resulting ternary electrolytes show both an increased conductivity and inherent stability. 
Moreover, ILs are non-volatile, non-flammable~\cite{AdamNature2000} and exhibit a wide 
electrochemical stability window~\cite{MacFarlaneNature1999}. 

However, it is not yet fully understood in how far the lithium transport mechanism in these 
materials changes relative to the conventional polymer electrolytes. 
For instance, it was speculated~\cite{PasseriniElectrochemCommun2003} that the lithium ions 
become progressively coordinated by the anions from the IL and are thus decoupled 
from the rather slow PEO chains. 
Alternatively, one might also expect that the IL enhances the PEO dynamics and serves as a 
plasticizer in this way, which is a common observation when adding low-molecular solvents to 
PEO-salt systems~\cite{BandaraElectrochimActa1998,BorghiniElectrochimActa1996,KimSSI2002}. 
In this work, we utilize molecular dynamics (MD) simulations to unravel the impact of the 
addition of IL. 
In order to quantify the lithium motion, we employ an analytical cation transport 
model~\cite{MaitraHeuerPRL2007,DiddensMamol2010}, which has originally been devised 
for binary polymer electrolytes. 

\begin{figure}
 \centering
 \includegraphics[scale=0.45]{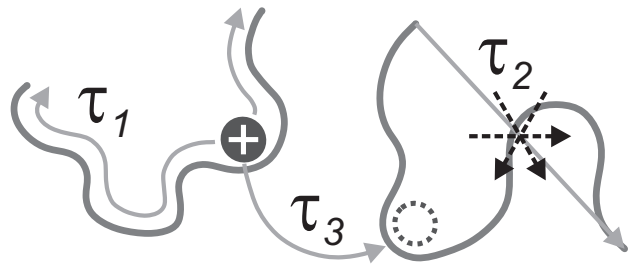}
 \caption{Scheme illustrating the three cation transport mechanisms in PEO-salt electrolytes. }
 \label{fig:sketch_transport_processes}
\end{figure}

Our description is based on both the Rouse model~\cite{RouseJCP1953} 
as well as the Dynamic Bond Percolation (DBP) model~\cite{rev_DBP}, 
and distinguishes three different microscopic lithium ion transport mechanisms 
(Figure~\ref{fig:sketch_transport_processes}): 
1.~The ions diffuse along the PEO backbone to which they are attached. 
This motion can be characterized by the time scale $\tau_1$ the ions need to explore the entire 
PEO chain. 
2.~For ambient temperatures, the PEO chains are naturally also subject to thermal motion, 
carrying the attached ions in this way. 
In case of Rousean motion, the polymer dynamics and thus motion of the attached ions can 
be quantified by an effective Rouse time $\tau_2$. 
3.~Finally, an ion bound to a specific PEO chain can be transferred to another chain. 
The mean residence time at a given chain is denoted as $\tau_3$ in the following. 
As demonstrated earlier~\cite{MaitraHeuerPRL2007}, the last mechanism can also be viewed 
as a renewal process within the framework of the DBP model. 

Of course, for the ternary electrolytes, it is a priori unclear if this scenario changes 
only quantitatively -- reflected by different values for $\tau_1$, $\tau_2$ and $\tau_3$ -- 
or if the lithium ion transport mechanism also changes on a qualitative level. 
In particular, we focus on two ternary polymer electrolytes with the same IL as in 
ref.~\citenum{PasseriniElectrochemCommun2003}, 
\ie {\it N}-methyl-{\it N}-propyl\-pyr\-rolid\-inium TFSI (PYR$_\mathrm{13}$TFSI), with a 
stoichiometry of \PsomeIL\ and \PmoreIL, respectively. 
The binary polymer electrolyte, \Ptwenty, serves as a reference. 
For convenience, PEO will be abbreviated as \lq P\rq\ and LiTFSI as \lq S\rq\ in the following, 
leading to the short-hand notation \Px\ with $x=0$, $x=0.66$ and $x=3.24$.

\section{Simulation Details}

The simulations were performed with the AMBER~10 package~\cite{AMBER10}. 
Here, the sander module was modified, allowing us to use a many-body polarizable force field 
specifically designed for PEO/LiTFSI~\cite{BorodinJPCB2006_PEO,BorodinJPCB2006_PEOLiTFSI} and 
PYR$_\mathrm{13}$TFSI~\cite{BorodinJPCB2006_IL}. 
The simulation box contained $10$ PEO chains with $N=54$ monomers each as well as $27$ LiTFSI 
ion pairs, yielding a concentration of ether oxygens (EOs) to lithium ions of $20:1$. 
Additionally, the two ternary systems contained $18$ or $87$ PYR$_\mathrm{13}$TFSI molecules, corresponding 
to $x=0.66$ and $x=3.24$. 
The simulation cells have been created randomly in the gas phase to yield homogeneous systems. 
After equilibration runs of $70-80\text{~ns}$ in the $NpT$ ensemble, production runs with a length 
of $200\text{~ns}$ have been performed in the $NVT$ ensemble, collecting data every picosecond. 
An elementary integration step of $1\text{~fs}$ was used, while the systems were coupled to 
a Berendsen thermostat~\cite{BerendsenJCP1984} with a reference temperature of $423\text{~K}$. 
All bonds involving hydrogen were constrained by the SHAKE algorithm~\cite{SHAKE}. 
The induceable point dipoles were integrated by a Car-Parrinello-like scheme~\cite{vanBelleMolPhys1992}. 
By comparing various radial distribution functions and mean square displacements (MSDs) for 
the first and the second half of the runs, we confirmed that the systems are in equilibrium. 
Moreover, the former showed no long-range ordering, demonstrating that the systems are perfectly 
mixed.

\section{Structural Properties}

\subsection{Lithium Ion Coordination}

We find for all electrolytes that virtually all lithium ions are coordinated to one or two 
PEO chains, thereby giving a first hint that also for the ternary systems the cation transport 
entirely takes place at the PEO chains. 
The percentages of lithium ions coordinating to one or two PEO chains is given in Table~\ref{tab:1PEO_2PEO_ratio}. 
With increasing IL concentration, the probability that a lithium ion coordinates to two PEO chains 
decreases, thus indicating a dilution effect which reduces the probability for an ion to encounter a second 
PEO molecule. 
This is also supported by the observation that the amount of \Li\ coordinating two PEO chains decreases 
linearly with $x$. 

\begin{table}
 \centering
 \begin{tabular}{l c c}
  \hline
  \hline
  system & 1 PEO [\%] & 2 PEO [\%] \\
  \hline
  \Ptwen\ & $47.2$ & $52.7$ \\
  \Psome\ & $53.0$ & $47.0$ \\
  \Pmore\ & $75.8$ & $24.0$ \\
  \hline
  \hline
 \end{tabular}
 \caption{Ratio of lithium ions coordinating to one or two PEO chains. }
 \label{tab:1PEO_2PEO_ratio}
\end{table}

\begin{figure}
 \centering
 \includegraphics[scale=0.3]{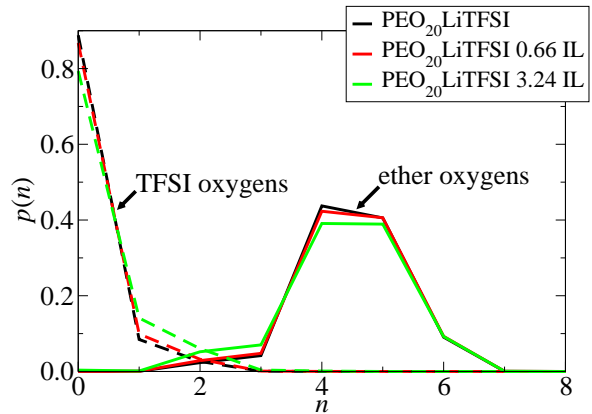}
 \caption{Probability $p(n)$ to find a certain coordination number $n$ of EOs 
          (irrespective if the ion is tied to one or two PEO chains) or TFSI oxygens. }
 \label{fig:coordbin}
\end{figure}

The predominant lithium coordination consists of $4-5$ EOs (see Figure~\ref{fig:coordbin}, which shows the probability distribution 
functions $p(n)$ to find a lithium ion with $n$ EOs or TFSI oxygens in its first coordination shell), which is in good 
agreement with experimental data~\cite{MaoPRL2000} and quantum chemistry calculations~\cite{JohanssonPolym1999,BaboulJAmChemSoc1999}. 
In those complexes where the $4-5$ EOs originate from a single PEO molecule, the polymer chain 
wraps helically around the cation. For complexes involving two PEO chains, typically $2-3$ EOs 
from each chain coordinate to the ion. 
Additional coordinations by TFSI oxygens are rather rare (about $12-20\text{~\%}$, see 
Figure~\ref{fig:coordbin}), and in most cases the anion coordinates only briefly to the lithium ion. 
Again, the amount of \Li\ coordinating to TFSI increases linearly with $x$, thus indicating 
that this effect is purely statistical. 

\subsection{Statical Polymer Properties}

For the conformational properties of the PEO molecules, we naturally observe a 
contraction of the chains due to the crown-ether-like coordination sphere of the 
lithium ions~\cite{MaitraHeuerMCP2007}. 
This manifests itself by a decrease of the mean squared end-to-end vector $\langle{\bf R}_\mathrm{e}^2\rangle$ 
for the electrolytes ($\langle{\bf R}_\mathrm{e}^2\rangle=1662\text{~\AA$^2$}$, $1570\text{~\AA$^2$}$ 
and $1571\text{~\AA$^2$}$ for $x=0$, $0.66$ and $3.24$, respectively) as compared to the pure PEO melt 
($\langle{\bf R}_\mathrm{e}^2\rangle=1979\text{~\AA$^2$}$). 
In case of the ternary electrolytes, one might expect that the PEO chains are swollen on a 
global scale, since the addition of IL would induce a crossover from a polymer melt to 
a semidilute solution. 
However, from both $\langle{\bf R}_\mathrm{e}^2\rangle$ and the scaling of the Rouse-mode 
amplitudes~\cite{RouseJCP1953,DoiEdwards} (not shown), we observe no such feature.

\section{Dynamics of the Lithium Ions}

\begin{figure}
 \centering
 \includegraphics[scale=0.3]{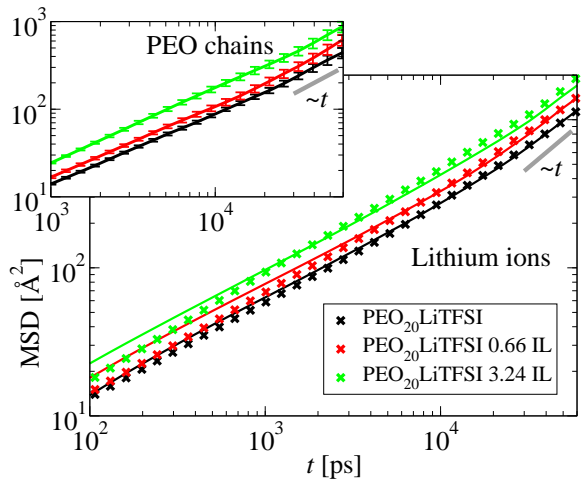}
 \caption{MSD of the lithium ions (main panel) and the center of mass of the PEO chains (inset). 
          The solid lines in the main panel correspond to the model predictions. }
 \label{fig:lithium_msd}
\end{figure}

Although from a structural point of view no significant differences emerge, we observe a clear increase 
of the lithium MSD, especially for the subdiffusive regime at $t=1-10\text{~ns}$ (crosses in Figure~\ref{fig:lithium_msd}), 
whereas the onset to diffusion occurs on comparable time scales, \ie $t=20-50\text{~ns}$. 
A similar increase can be found for the MSD of the entire PEO chains (inset of Figure~\ref{fig:lithium_msd}). 
In the following, we will go more into detail and investigate the relative importance of the individual 
transport mechanisms.

\subsection{Interchain Transfer}

In order to calculate the renewal times, the number of transfer processes $N_\mathrm{tr}$ 
was counted from the simulations, and the $\tau_3$-values were determined according to 
$\tau_3=t_\mathrm{max}N_{\mathrm{Li}^+}/N_\mathrm{tr}$, where $t_\mathrm{max}=200\text{~ns}$ 
is the simulation length and $N_{\mathrm{Li}^+}=27$ is the number of lithium ions in the 
simulation box. 
Of course, it is questionable if brief transfers followed by successive backjumps to the 
previous polymer chain serve as a renewal process in the strict sense, since the lithium 
dynamics will not become uncorrelated to its past after such an event. 
A more detailed analysis (not shown) revealed that these non-Markovian, short-time backjumps 
occurred up to $100\text{~ps}$, which we used subsequently as a criterion to define real 
renewal events. 
In cases where the transfer was mediated by TFSI anions only (probability $p_\mathrm{IL}$ in 
Table~\ref{tab:taus}), we found that the displacement the ion covers in the IL-rich region was 
sufficiently small, so that the contribution of these transfers to the lithium MSD is 
negligible. 

\begin{table*}
  \centering
  \begin{tabular}{c c c c c c c c c c}
    \hline
    \hline
    $x_\mathrm{IL}$ & $\tau_1\text{~[ns]}$ & $\langle{\bf R}_\mathrm{e}^2\rangle\text{~[\AA$^2$]}$ & $\tau_\mathrm{R}\text{~[ns]}$ & $\tau_2\text{~[ns]}$ & $\tau_3\text{~[ns]}$ & $p_\mathrm{IL}\text{~[\%]}$ & $D_\mathrm{Li}^\mathrm{sim}\text{~[\AA$^2$ns$^{-1}$]}$ & $D_\mathrm{Li}^{\infty}\text{~[\AA$^2$ns$^{-1}$]}$ & $D_\mathrm{Li}^\mathrm{exp}\text{~[\AA$^2$ns$^{-1}$]}$~\cite{PasseriniJoost} \\  
    \hline
    $0.0$ & $147$ & $1662$ & $45$ & $167$ & $17$ & $2.5$ & $2.945$ & $1.947$ & $0.052$ ($x=0.0$) \\
    $0.66$ & $140$ & $1570$ & $37$ & $89$ & $18$ & $1.0$ & $3.542$ & $2.309$ & $0.118$ ($x=1.0$) \\
    $3.24$ & $127$ & $1571$ & $24$ & $68$ & $24$ & $8.5$ & $4.257$ & $2.392$ & $0.126$ ($x=4.0$) \\
    \hline
    \hline
  \end{tabular}
  \caption{Parameters characterizing the three transport mechanisms (see text for further explanation). }
  \label{tab:taus}
\end{table*}

We observe that $\tau_3$ (Table~\ref{tab:taus}) increases with increasing IL concentration. 
Since the PEO molecules become more and more diluted, this can mainly be explained as a 
concentration effect. 
Obviously, the critical step for a transfer process is the encounter of a another PEO 
segment.

\subsection{Motion Along the PEO Backbone}

\begin{figure}
 \centering
 \includegraphics[scale=0.3]{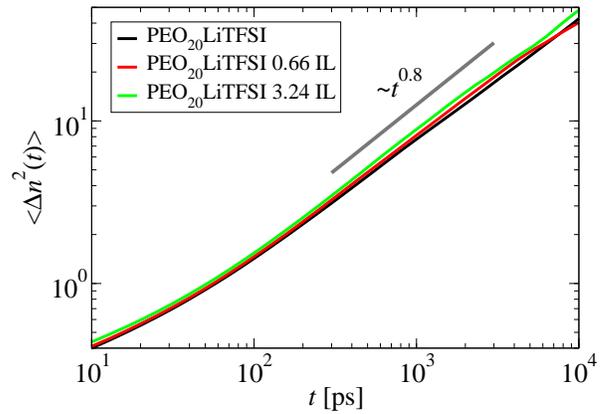}
 \caption{MSD of the average EO index $\langle\Delta n^2(t)\rangle$. }
 \label{fig:delta_n_msd}
\end{figure}

In order to quantify the diffusion along the PEO backbone, we successively numbered all 
monomers at a given PEO chain, allowing us to express the lithium position by the average 
EO index $n$, and to calculate an effective MSD $\langle\Delta n^2(t)\rangle$ along this 
coordinate (Figure~\ref{fig:delta_n_msd}). 
We find that this type of motion is slightly subdiffusive (\ie $\langle\Delta n^2(t)\rangle\propto t^{0.8}$) 
for all electrolytes within the statistical error. 
No significant dependence on the IL concentration can be observed. 
This indicates that the surrounding molecules (PEO chains or IL) have no influence on this 
mechanism. 
Moreover, also the magnitude of $\langle\Delta n^2(t)\rangle$ is essentially the same 
for lithium ions bound to one or to two PEO chains (not shown). 

Note that the statistics of the $\langle\Delta n^2(t)\rangle$-curves in Figure~\ref{fig:delta_n_msd} are 
insufficient for $t>10\text{~ns}$, wherefore the plot is only shown up to $10\text{~ns}$. 
This is due to the fact that the lithium ions are transferred to other PEO chains, and the 
motion cannot be tracked any further in a reliable manner, even though in some cases the ion 
jumps back to the first chain after some picoseconds. 
For sufficiently long chains as well as a significant amount of $\text{PEO}-$\Li\ complexes 
that exist throughout the entire observation time, $\langle\Delta n^2(t)\rangle$ will 
show diffusive behavior on longer time scales. 
Naturally, in the limit $t\rightarrow\infty$, one would expect a crossover to a plateau for 
finite chain lengths. 
From $\langle\Delta n^2(t)\rangle$, neither of these two effects can be found, indicating that most 
life times of the $\text{PEO}-$\Li\ complexes are too short. 
Rather, the lithium ions only move on average $7-8$ monomers during $10\text{~ns}$. 
Keeping in mind that the lithium ions are typically bound to $4-5$ monomers, these findings imply 
that the ions have barely left their own coordination sphere during the accessible time scale. 
Therefore, finite size effects of the PEO chains are irrelevant in the present case. 

In principle, the motion along the PEO chain can be quantified by an effective diffusion coefficient 
$D_1$, which can be calculated from the $\langle\Delta n^2(t)\rangle$-curves according to 
\begin{equation}
 D_1=\frac{\langle\Delta n^2(t)\rangle}{2\,t}\,\text{.}
\end{equation}
However, due to the subdiffusivity of $\langle\Delta n^2(t)\rangle$, the $D_1$ values depend on the specific 
time for which they are estimated. 
Thus, in order to estimate the net effect of this mechanism, one would ideally compute $D_1$ at $t=\tau_3$, 
for which the statistics are unfortunately too bad due to reasons mentioned above. 
As an approximation, we extrapolated the $\langle\Delta n^2(t)\rangle$-curves in Figure~\ref{fig:delta_n_msd} under the 
assumption that the scaling $\langle\Delta n^2(t)\rangle\propto t^{0.8}$ persists until $t=\tau_3$. 
In order to estimate the net effect of this mechanism (\ie the number of traversed monomers 
before the ion is transferred to another chain), we define $\tau_1$ via~\cite{MaitraHeuerPRL2007}
\begin{equation}
 \tau_1=\frac{(N-1)^2}{\pi^2}\frac{2\,\tau_3}{\langle\Delta n^2(\tau_3)\rangle}\,\text{,}
\end{equation}
which due to the subdiffusivity of $\langle\Delta n^2(t)\rangle$ slightly depends on $\tau_3$. 
One observes that $\tau_1$ decreases slightly with increasing IL concentration (Table~\ref{tab:taus}), 
reflecting the weak dependence of $\tau_1$ on $\tau_3$.

\subsection{Polymer Motion}

\begin{figure}
 \centering
 \includegraphics[scale=0.3]{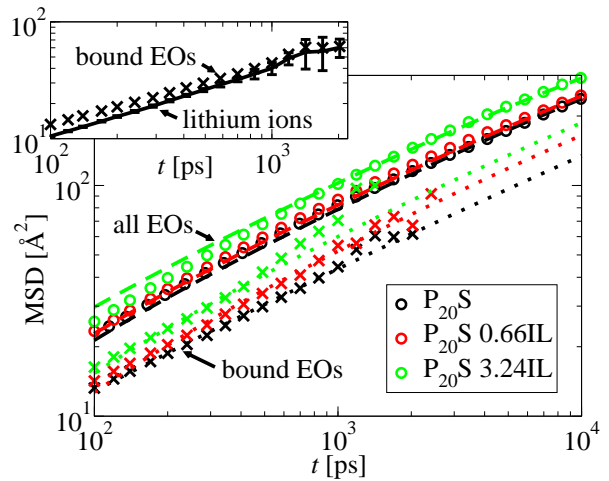}
 \caption{MSDs of the average EOs (circles), the bound EOs (crosses) and the lithium ions 
          bound to these EOs (inset, solid lines). 
          The dashed and dotted lines show the respective Rouse fits. }
 \label{fig:EO_Li_msd}
\end{figure}

Figure~\ref{fig:EO_Li_msd} shows the MSD of the EOs relative to the center of mass of the PEO chain. 
This quantity has been computed for all EOs (\ie irrespective of the presence of an ion), for 
EOs bound to a lithium ion as well as for the respective attached ions. 
The criterion to consider a cation or EO as bound was that the average EO index of the ion 
did not change more than one, \ie $\left|\Delta n(t)\right|\leq 1$ for all time frames 
during $t$. 
For the bound EOs, no further distinction between additional coordinations of the lithium ion 
to another PEO chain or a TFSI molecule was made. 
Thus, these effects are already implicitly contained in the curves in Figure~\ref{fig:EO_Li_msd}. 
Of course, it is questionable if cations bound to two PEO chains show the same dynamics as 
ions bound to one chain only, since the former could be regarded as transient crosslinks, 
which would significantly impede the polymer motion. 
A more detailed analysis indeed revealed that there is a conceptual difference between 
these two coordinations, however, this effect can easily be taken into account (see Appendix~\ref{app:xlinks}) 
and does not affect the general formalism of our analysis. 

The average EOs (circles) show typical Rouse-like motion with the characteristic relaxation 
time $\tau_\mathrm{R}$. 
The dynamics of the bound EOs (crosses) is qualitatively the same but protracted. 
Therefore, it is possible to characterize the dynamics of the bound EOs by a larger, effective 
Rouse time $\tau_2$. 
The lithium ions attached to these EOs (shown in the inset of Figure~\ref{fig:EO_Li_msd} for \Ptwen, 
the curves for the other electrolytes look similar) closely follow the bound EOs, which gives 
clear evidence for their cooperative motion. 
On short time scales, the MSD of the EOs is larger than the lithium MSD due to the additional 
internal degrees of freedom of the PEO backbone, but the MSD of the bound cations catches up at 
$t\approx 1\text{~ns}$. 
Thus, $\tau_2$ characterizes both the dynamics of the bound PEO segments as well as of the 
attached lithium ions. 

Figure~\ref{fig:EO_Li_msd} also shows the Rouse fits, \ie 
$g_\mathrm{R}(t) = 2\langle{\bf R}_\mathrm{e}^2\rangle\,\pi^{-2}\sum_{p=1}^{N-1}\left[1-\exp{(-tp^2/\tau_\mathrm{R})}\right]\,p^{-2}$, 
for the average (dashed lines) and for the bound EOs (dotted lines). 
Of course, the precise value of $\tau_\mathrm{R}$ and $\tau_2$ also depends on the 
value of $\langle{\bf R}_\mathrm{e}^2\rangle$. 
In order to obtain a fit consistent with the plateau value at large $t$ (not shown in Figure~\ref{fig:EO_Li_msd} for clarity), 
the MSDs of the average EOs were fitted using two parameters, \ie $\tau_\mathrm{R}$ and 
$\langle{\bf R}_\mathrm{e}^2\rangle$. 
Subsequently, the MSDs of the bound EOs were fitted using this value in combination with a 
single fit parameter $\tau_2$ only 
(Table~\ref{tab:taus}, deviations from our previous study~\cite{DiddensMamol2010} on \Ptwen\ 
arise from the shorter simulation length of about $27\text{~ns}$ and the modified fitting 
procedure). 

Whereas the $\langle{\bf R}_\mathrm{e}^2\rangle$-values are approximately constant, 
both $\tau_\mathrm{R}$ and $\tau_2$ decrease significantly, clearly indicating 
that the dynamics of the PEO segments becomes faster with increasing IL concentration. 
Therefore, the IL can be regarded as plasticizer. 
For the average segments, the dynamics for \Pmore\ is nearly the same as for pure PEO 
($\tau_\mathrm{R}=22\text{~ns}$), showing that the plasticizing approximately cancels 
with the slowing-down caused by the coordinating lithium ions as found for \Ptwen. 
The presence of the IL also enhances the motion of the bound segments, and, as a result, 
the dynamics of the respective attached lithium ions, leading to an increase of the 
overall lithium MSD. 
Here, experimental studies reported similar findings for other plasticizers like 
ethylene/propylene carbonate~\cite{BandaraElectrochimActa1998,KimSSI2002} or short PEO 
chains embedded in a high-molecular weight matrix~\cite{BorghiniElectrochimActa1996,KimSSI2002}. 

For finite $N$, the plasticizing effect is even twofold. 
Apart from the internal, segmental PEO dynamics (Figure~\ref{fig:EO_Li_msd}), the center-of-mass 
motion is also accelerated by the addition of IL (inset of Figure~\ref{fig:lithium_msd}). 
The relative importance of these two types of plasticizing will be discussed below.

\section{Application of the Transport Model}

As a consistency check of our description, we employ the transport model to reproduce 
the lithium MSD in Figure~\ref{fig:lithium_msd}. 
During the residence time $\tilde{t}$ at a given PEO chain, the MSD $g_{12}$ of the lithium ion 
is given by a Rouse-like expression~\cite{MaitraHeuerPRL2007} 
\begin{equation}
 \label{eq:g12model}
 g_{12}(\tilde{t}) = \frac{2\langle{\bf R}_\mathrm{e}^2\rangle}{\pi^2}\sum_{p=1}^{N-1}\frac{\left[1-\exp{\left(-\frac{\tilde{t}p^2}{\tau_{12}}\right)}\right]}{p^2}\,\text{,}
\end{equation}
where $\tau_{12}^{-1}=\tau_1^{-1}+\tau_2^{-1}$ is a combined relaxation rate due to both intramolecular 
mechanisms. 
After a renewal process (\ie an interchain transfer), the ion dynamics becomes uncorrelated to 
its past~\cite{rev_DBP,MaitraHeuerPRL2007}, and the motion at the new chain is again characterized 
by Eq.~\ref{eq:g12model}. 
Thus, it is possible to interpret the overall lithium dynamics as a random walk, in which the 
elementary step length is given by Eq.~\ref{eq:g12model}, and the direction of motion for the successive 
step changes randomly after each renewal event. 
We assumed that the number of ion transfer processes during time interval $t$ is given by a Poisson 
distribution with mean $t/\tau_3$, leading to exponentially distributed $\tilde{t}$. 
For a given $t$, the lithium MSD due to the three transport mechanisms was then obtained from the 
numerical average over a large number of Poisson processes. 
The third ingredient required for the total lithium MSD is the center-of-mass motion of the PEO chains, 
which was directly extracted from the simulations (inset of Figure~\ref{fig:lithium_msd}) and added to the 
model curve. 
The resulting predictions are shown as solid lines in the main panel of Figure~\ref{fig:lithium_msd}, the 
respective diffusion coefficients $D_\mathrm{Li}^\mathrm{sim}$ calculated from the model curves are 
given in Table~\ref{tab:taus}. 

For \Ptwen, one observes a nearly perfect agreement throughout the entire observation time. 
This demonstrates that our transport model indeed captures the underlying, much more complex 
microscopic scenario. 
In case of \Psome\ and \Pmore, the model curves agree with the empirical lithium MSD for time 
scales larger than about $1-2\text{~ns}$. 
Slight deviations can be attributed to the large uncertainties of the MSD of the PEO chains. 
However, the model prediction systematically overestimates the MSDs of \Psome\ and \Pmore\ 
for short time scales. 
Here, a more detailed analysis (to be published under separate cover) revealed that these deviations 
are caused by hydrodynamic interactions arising from the presence of the IL. 
On larger length and time scales, these hydrodynamic interactions are screened, which has also been 
reported for other semidilute polymer solutions~\cite{DuenwegPRE2001}. 
Thus, both the Rouse-like behavior and the diffusive regime are correctly reproduced, which clearly 
demonstrates the applicability of our model to the experimentally relevant long-time limit. 

Finally, we use the same procedure as above to compute $D_\mathrm{Li}^\infty$ for $N\rightarrow\infty$ 
via the scaling laws~\cite{MaitraHeuerPRL2007} $\langle{\bf R}_\mathrm{e}^2\rangle\propto N$, 
$\tau_1\propto N^2$, $\tau_2\propto N^2$ and $\tau_3\propto N^0$. 
Of course, for the scaling of $\tau_2$, entanglement effects may become relevant, which would slow down 
the segmental dynamics. 
However, if $\tau_3<\tau_\mathrm{e}$ (\ie the entanglement time), meaning that the lithium ion leaves the 
PEO chain before the latter begins to reptate, the overall dynamics is still Rousean~\cite{DoiEdwards}, 
and our model can still be used to calculate $D_\mathrm{Li}$. 
For PEO, experiments~\cite{ShiSSI1993} revealed that the entanglement regime sets in from about $N\approx 75$. 
Based on these observations, one can estimate $\tau_\mathrm{e}$ according to $\tau_\mathrm{e}=\tau_\mathrm{R}(N=75)=\tau_\mathrm{R}(75/54)^2$. 
For \Ptwen, this leads to $\tau_\mathrm{e}\approx 87\text{~ns}$, which is substantially larger than $\tau_3$. 
Also in case of the highly plasticized \Pmore\ one finds $\tau_\mathrm{e}\approx 46\text{~ns}>\tau_3$. 
Therefore, the lithium ion leaves the PEO chain before the tube constraints become noticeable, 
and our formalism can also be applied for $N\rightarrow\infty$. 

Table~\ref{tab:taus} shows $D_\mathrm{Li}^{\infty}$ calculated from the model together with 
the PFG-NMR data~\cite{PasseriniJoost} at $T=323\text{~K}$. 
For the experimental measurements, both the IL fraction $x$ and the IL cation, 
\ie PYR$_\mathrm{14}$, are slightly different than in our simulations, however, one 
would expect no significant effect on the transport mechanism. 
In both cases, we observe a clear increase of $D_\mathrm{Li}$, which can be attributed 
to the plasticizing effect of the IL. 

However, when discussing these values, one has to keep in mind that not only the segmental 
motion, but also $\tau_3$ and $D_\mathrm{PEO}$ affect the precise value of $D_\mathrm{Li}$, 
in which each contribution has its own $N$-dependence. 
For example, in case of $N\rightarrow\infty$, the mean intramolecular MSD $\langle g_{12}\rangle$ 
(averaged over all $\tilde{t}$, Eq.~\ref{eq:g12model}), increases by about $28\text{~\%}$ for \Psome\ 
and $73\text{~\%}$ for \Pmore, mainly as a result of the increased segmental mobility. 
On the other hand, the renewal rate decreases by about $7\text{~\%}$ and $29\text{~\%}$, 
although the plasticizing effect dominates, and the overall $D_\mathrm{Li}^{\infty}$-values 
increase by about $19\text{~\%}$ and $23\text{~\%}$. 
For $N=54$, the situation is slightly different. 
Here, the segmental plasticizing, measured by $\langle g_{12}\rangle$, leads only to an 
increase of $22\text{~\%}$ for $x=0.66$ and $54\text{~\%}$ for $x=3.24$. 
Finally, for $N\rightarrow 1$, the differences in $\langle g_{12}\rangle$ would even 
disappear~\cite{MaitraHeuerPRL2007}. 
However, this trend is compensated by the plasticizing of the center-of-mass motion 
of PEO. 
For $N=54$, $D_\mathrm{PEO}$ is raised by $30\text{~\%}$ for $x=0.66$ and by $92\text{~\%}$ 
for $x=3.24$, which results in an overall increase of $D_\mathrm{Li}$ of $20\text{~\%}$ for 
\Psome\ and $45\text{~\%}$ for \Pmore. 

So far, we focused on the high-temperature limit which we can address in our simulations. 
Interestingly, the relative increase of $D_\mathrm{Li}^\mathrm{exp}$ upon the addition of 
IL becomes much more pronounced in the low-temperature regime~\cite{PasseriniJoost} (see 
also Table~\ref{tab:taus}). 
Although simulations at low temperatures would be too costly, one might expect that the 
plasticizing effect at least partly accounts for the larger relative increase of $D_\mathrm{Li}$ 
in this regime. 
Here, DSC measurements~\cite{PasseriniJoost} revealed that the glass-transition temperature 
decreases up to $35\text{~K}$ upon IL addition, which gives a first hint that also at low 
temperatures the enhanced polymer dynamics contributes to the faster lithium motion. 
In such a scenario, the plasticizing of the polymer matrix via electrochemically stable 
additives would be an important milestone for the use of SPE-based batteries in electronic 
devices, as their current limitation particularly holds for low (\ie ambient) temperatures. 
In fact, PEO/LiTFSI/IL mixtures have recently been successfully applied in prototype 
batteries~\cite{BalducciJPowerSources2011}.

\section{Conclusion}

In this study, we have examined the microscopic lithium ion transport 
mechanism in ternary polymer electrolytes consisting of \Ptwenty\ and 
PYR$_\mathrm{13}$TFSI. 
In particular, we addressed the question in how far the microscopic 
scenario of the ion transport changes upon the addition of IL, and 
how the experimentally observed increase in the lithium ion diffusion 
coefficient~\cite{PasseriniJoost} can be understood in terms of the 
individual transport mechanisms. 
To this purpose, an analytical cation transport model~\cite{MaitraHeuerPRL2007} 
was successfully applied. 
It turned out that virtually all lithium ions were coordinated to the PEO 
chains, thus ruling out a transport mechanism in which the lithium transportation 
is decoupled from the polymer chains. 
Rather, the main reason for the increase of the lithium diffusion coefficient, at 
least for the considered temperature, is the plasticizing effect of the IL, which 
enhances the segmental motion of the PEO chains and thus also the dynamics of the 
attached ions. 
A minor counteracting effect was the successive dilution of the electrolyte due to 
the IL, which slightly decreases the rate of interchain transfers. 
In the sum however, the plasticizing is dominant, and the overall lithium diffusivity 
increases. 
For the design of novel battery materials, our findings therefore imply that 
a polymer electrolyte which is both highly plasticized and exhibits a high transfer 
rate, \eg facilitated by a more coordinating IL, would yield optimal results.

\begin{acknowledgments}
The authors would like to thank Oleg Borodin, Nicolaas A. Stolwijk, Stefano Passerini 
and Mario Joost for helpful discussions and for providing the experimental 
data. 
Financial support from the NRW Graduate School of Chemistry is also greatly 
appreciated. 
\end{acknowledgments}

\begin{appendix}

\section{Influence of the \Li\ Coordination on the Segmental Motion}
\label{app:xlinks}

\begin{figure}
 \centering
 \includegraphics[scale=0.3]{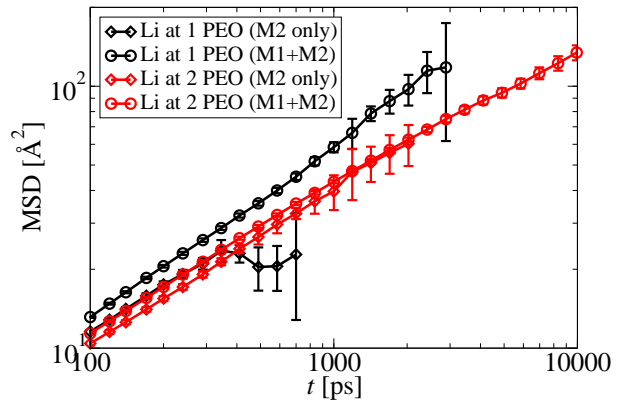}
 \caption{MSDs of the lithium ions bound to one or two PEO chains 
          in \Ptwenty. A second distinction was made between cations 
          that diffused along the PEO chain (M1 and M2) and 
          ions that remained bound to the same EOs (M2 only). 
          All curves have been computed in the center-of-mass frame. 
          For the other electrolytes, the scenario is qualitatively the same. }
 \label{fig:Li_msd_1PEO_vs_2PEO}
\end{figure}

In order to elucidate in how far the lithium ions coordinated to two PEO chains act as 
temporary crosslinks, Figure~\ref{fig:Li_msd_1PEO_vs_2PEO} shows the MSD of lithium ions 
bound to one and to two PEO chains for \Ptwen. 
A second distinction was made if the ions diffused along the PEO chain (\ie the ion was 
transported by both intrachain diffusion (M1) and polymer dynamics (M2)) or 
remained bound to the same EOs (polymer dynamics only, here, the same criterion as for Figure~\ref{fig:EO_Li_msd} 
of the main part of our analysis, $\left|\Delta n(t)\right|\leq 1$, has been applied). 
In case of cations bound to one PEO chain only, ions undergoing both types of intramolecular 
transport are clearly faster than those remaining close to the initial position on the PEO 
chain (Figure~\ref{fig:Li_msd_1PEO_vs_2PEO}). 
For lithium ions connected to two PEO chains, no significant difference between 
these two categories can be observed in the MSD (Figure~\ref{fig:Li_msd_1PEO_vs_2PEO}). 
This implies that the cations bound to two PEO chains experience no effective 
transport due to the diffusion along the chain. 
Rather, the PEO chain moves reptation-like along its own contour past the cation, 
which results in a non-zero $\langle\Delta n^2(t)\rangle$, but does not contribute 
to the overall lithium transport. 
For all other investigated electrolytes, the observations from Figure~\ref{fig:Li_msd_1PEO_vs_2PEO} 
are qualitatively the same. 

\begin{figure}
 \centering
 \includegraphics[scale=0.3]{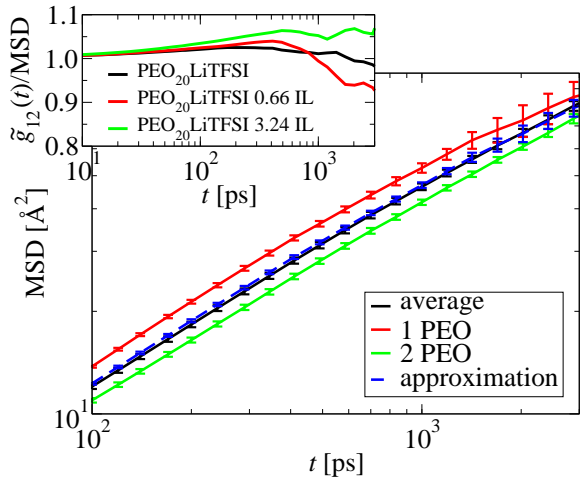}
 \caption{MSD of lithium ions bound to one or two PEO chains as well as 
          the MSD of cations which are bound to the same PEO molecule during $t$ 
          (irrespective of further coordinations) for \Ptwen. 
          The MSDs of ions bound to one and to two chains have been used to compute 
          an approximate average $\tilde{g}_{12}(t)$ according to Eq.~\ref{eq:g12_g2}. 
          Inset: ratio of $\tilde{g}_{12}(t)$ relative to the real average MSD as 
          extracted from the simulations. }
 \label{fig:1PEO_vs_2PEO_approx}
\end{figure}

With respect to the transport model, in particular Eq.~\ref{eq:g12model} in the main body, this 
effect can be easily captured. 
Here, the only additionally required parameter is the ratio of lithium ions 
coordinated to one or two PEO chains (Table~\ref{tab:1PEO_2PEO_ratio}): 
\begin{equation}
 \tilde{g}_{12}(t) = r_\mathrm{1\,PEO}\,g_{12}(t) + (1 - r_\mathrm{1\,PEO})\,g_{2}(t) 
 \label{eq:g12_g2}
\end{equation}
For the fraction $r_\mathrm{1\,PEO}$ of cations bound to one PEO chain, 
Eq.~\ref{eq:g12model} remains valid, whereas for ions bound to 
two chains with $\tau_{1,\mathrm{\,2\,PEO}}\rightarrow\infty$ only $\tau_2$ 
is important. 
Within this approximation, the average dynamics of the lithium ions (\ie averaged over ions 
bound to one or two PEO chains while simultaneously allowing the intermediate exchange between 
these two coordination states) is estimated from the structural property $r_\mathrm{1\,PEO}$ only. 

Figure~\ref{fig:1PEO_vs_2PEO_approx} shows that this rather simplistic picture is indeed valid 
to a good approximation. 
Here, Figure~\ref{fig:1PEO_vs_2PEO_approx} displays the MSD of ions bound to one or two PEO 
molecules in \Ptwen\ as extracted from the simulations. 
These curves have in turn been used to calculate an approximate average MSD according 
to Eq.~\ref{eq:g12_g2}, which is also shown in Figure~\ref{fig:1PEO_vs_2PEO_approx}. 
In fact, the agreement of $\tilde{g}_{12}(t)$ with the average MSD (\ie a lithium ion 
that remained on the same chain irrespective of other coordinations) is nearly 
quantitative. 
Deviations for larger time scales are due to bad statistics. 
In the same spirit, the inset of Figure~\ref{fig:1PEO_vs_2PEO_approx} shows the ratio between the 
approximate $\tilde{g}_{12}(t)$ and the average MSD directly calculated from the simulations 
for all systems. 
As for \Ptwen, this ratio is close to unity for all other electrolytes. 
These observations highlight that, since the crosslinks are temporary and an exchange 
between both coordination types takes place, the long-time behavior of the 
intramolecular dynamics may be estimated by the average in Eq.~\ref{eq:g12_g2}, 
as nicely demonstrated in Figure~\ref{fig:1PEO_vs_2PEO_approx}. 
\end{appendix}

\bibliography{thebib}

\end{document}